\documentclass[conference,9pt]{IEEEtran}

\usepackage[pdftex]{graphicx}
\usepackage{amsmath}
\usepackage{amssymb}

\usepackage{color}

\IEEEoverridecommandlockouts

\newcommand{\bv}{\boldsymbol{v}}
\newcommand{\bn}{\boldsymbol{n}}
\newcommand{\br}{\boldsymbol{r}}
\newcommand{\be}{\boldsymbol{e}}
\newcommand{\by}{\boldsymbol{y}}
\newcommand{\bx}{\boldsymbol{x}}
\newcommand{\bp}{\boldsymbol{p}}
\newcommand{\bu}{\boldsymbol{u}}
\newcommand{\bz}{\boldsymbol{z}}

\newcommand{\btheta}{\boldsymbol{\theta}}

\newcommand{\bW}{\boldsymbol{W}}
\newcommand{\bT}{\boldsymbol{T}}
\newcommand{\bF}{\boldsymbol{F}}
\newcommand{\bI}{\boldsymbol{I}}
\newcommand{\bH}{\boldsymbol{H}}
\newcommand{\bG}{\boldsymbol{G}}

\begin{document}

\title{Distributed image reconstruction for \\very large arrays in radio astronomy}

\author{\IEEEauthorblockN{Andr\'e Ferrari, David Mary, R\'emi Flamary and C\'edric Richard}
\IEEEauthorblockA{
Laboratoire Joseph-Louis Lagrange\\
Universit\'e de Nice Sophia-Antipolis, CNRS, Observatoire de la C\^ote d'Azur\\
Nice, France\\
Email: surname.name@unice.fr}
\thanks{This work was supported by CNRS grant MASTODONS; DISPLAY project.}}

\maketitle

\begin{abstract}
Current and future radio interferometric arrays such as LOFAR and SKA are characterized 
by a paradox. Their large number of receptors  (up to millions) allow theoretically unprecedented 
high imaging resolution. In the same time, the ultra massive amounts of samples makes 
the data transfer and computational loads (correlation and calibration)
order of magnitudes too high to allow any currently existing image reconstruction 
algorithm to achieve, or even approach, the theoretical resolution. We investigate 
here decentralized and distributed image reconstruction strategies which select, transfer and process 
only a fraction of the total data. The loss in MSE incurred by the proposed approach is evaluated 
theoretically and numerically on simple test cases.
\end{abstract}

\section{Introduction}
\textbf{}
Since the commissioning of the first large radio interferometers in
the 70s and 80s (such as the VLA in the USA and the WSRT) radio
astronomy in the range of large wavelengths has grown dramatically,
particularly with the development of more and more extended antenna arrays. 
In the prospect of the most sensitive radio telescope ever built, the SKA which will be operational in the 2020s, several new generation radio telescopes are being built or planned (LOFAR in the Netherlands, ASKAP and the Murchison Widefield Array Australia, e-MERLIN in the UK, e-EVN based in Europe, MeerKAT in South Africa, JVLA the United States).

As an example, LOFAR consists of 48 groups of antennas (stations),
among which approximately 35,000 elementary antennas are located in
the Netherlands. The ``superterp'', the heart of LOFAR is a super-station: a cluster of six stations. 
Eight other stations, totalizing approximately 13,000
antennas are located in the surrounding countries. A project of a new super-station
in Nan\c{c}ay (F) is under consideration. Within each station,
antennas form a phased array which allows for digital beamforming simultaneously in several directions
and frequency bands.
The beam-formed data from the stations are centralized at the University of Groningen in the
Netherlands where a supercomputer is responsible for the  combination
of the beam data from all stations.
The resulting data are then stored on a cluster of ASTRON, the Netherlands
Institute for Radio Astronomy, where the images (and other
deliverables) are reconstructed. As a mean of comparison SKA will
totalize 2.5 millions antennas, with a square kilometer collecting area
distributed over an area of $\approx$ 5,000 km diameter.

Beyond specific objectives that distinguish  these new fully digital ``software
telescopes'', they are all characterized by a great flexibility. Another common point is the
amount of data
which must be transferred to the central computer and processed.
It amounts to 1~terabit/second for LOFAR and will be of the order of $14$~exabyte/day for SKA
(more than 100 times the global internet traffic).  LOFAR uses a  1.5 Blue Gene/P 
for the data reduction and the computation of correlations. IBM et ASTRON will develop by 2024 a supercomputer to
process and store 1 petabytes of data everyday \cite{Broekema:2012cz}.

This correspondence investigates the possibility to distribute 
the image reconstruction over the super-stations. The main objective is to avoid centralization
of  the sampled electromagnetic fields acquired by all stations in order to reduce the data transfer and the exponential increase in the calibration and computational load.

Section II recalls the basis of radio astronomy with aperture synthesis and
proposes a strategy where each super-station uses all its antenna and one reference signal from the other 
super-stations.
The loss of performances that follows is evaluated on a simple model using the Cram\'er Rao Lower Bound (CRLB).
Section III shows that the image reconstruction problem can be written as a
global variable consensus problem with regularization. Numerical simulations illustrate the 
performances of the proposed approach. A concluding section presents perpectives.

\section{Aperture synthesis for radio astronomy}

\subsection{Standard aperture synthesis model}
This section provides the basic equations of radio astronomy with multiple sensor array
and describes a partial aperture synthesis strategy which aims to reduce data transfer, allowing
a decentralized image reconstruction.

To simplify the notations and without loss of generality, we will not make explicit the wavelengths  dependence
and the Earth rotation and assume punctual antennas. The coordinates of the stations 
(within each station, a beam is created from the phased array) in a plane perpendicular to the line of sight
are denoted as $\br_j$
and the map of interest (the ``image'' of a region of the sky) is  $x(\bp)$ where $\bp$  denotes the 
angular coordinates on the sky.
The fundamental equation of interferometry relates the Fourier transform of the map to the spatial coherency (visibility)
of the incoming electromagnetic field. A measurement of the coherency is obtained by correlating the signal acquired 
by a pair of stations $(i,j)$  
properly delayed located at $\br_i$ and $\br_j$, giving in the noiseless case a point of visibility at spatial frequency $\bu_\ell = \br_j-\br_i$:
\begin{equation}
v(\bu_\ell) = \int x(\bp) e^{-\jmath 2\pi \bu_\ell^t \bp }d\bp 
\end{equation}
See for example \cite{Thompson:2008ww,vanderVeen:2013uz} for a comprehensive
description of radio astronomy and signal processing related tools.

Computation of $v(\bu_\ell)$ obviously requires the transfer of signals from stations $i$ and $j$
in the same place. The stations are normally grouped in ``super-stations'' (e.g. the superterp for LOFAR) 
accounting  for low frequencies ($\|\br_i-\br_j\|_2$ small).
Resolution is then increased by correlating signals between stations that can be located up to thousands
of kilometers from each other.

\subsection{Reduced synthesis aperture model}

In general, all possible correlations are computed in order to maximize the 
$(u,v)$ coverage (set of $\bu_l$).
This requires a centralized system architecture. The resulting visibilities measurements 
$v(\bu_\ell)$ are associated to the
filtering of the visibility function $v(\bu)$ by the Full Aperture (FA)  spatial
transfer function
\begin{align}
\mathcal{A}_F(\bu)&=\left(\sum_{k} \delta(\br-\br_k)\ast\sum_{l} \delta(\br-\br_l)\right)(\bu)\\
&=\sum_{\ell=1}^M w_\ell \delta(\bu-\bu_\ell)=\mathcal{A}_L(\bu)+\mathcal{A}_H^F(\bu) \label{fullcover}
\end{align}
where $\ast$ denotes the  convolution, the weights 
$w_\ell$ count  the number of beam pairs measuring
the same spatial frequency $\bu_\ell$ and $M$ is  the number of different sampled frequencies. 
The term $\mathcal{A}_L(\bu)$ is associated to ``intra-super-stations'' low-frequency correlations
and $\mathcal{A}_H^F(\bu)$ to high-frequency ``inter-super-stations'' correlations. 
Note that in order to simplify the derivations, we will  also denote super-station
a single remote station.

\begin{figure}
\centering{\includegraphics[width=.75\columnwidth]{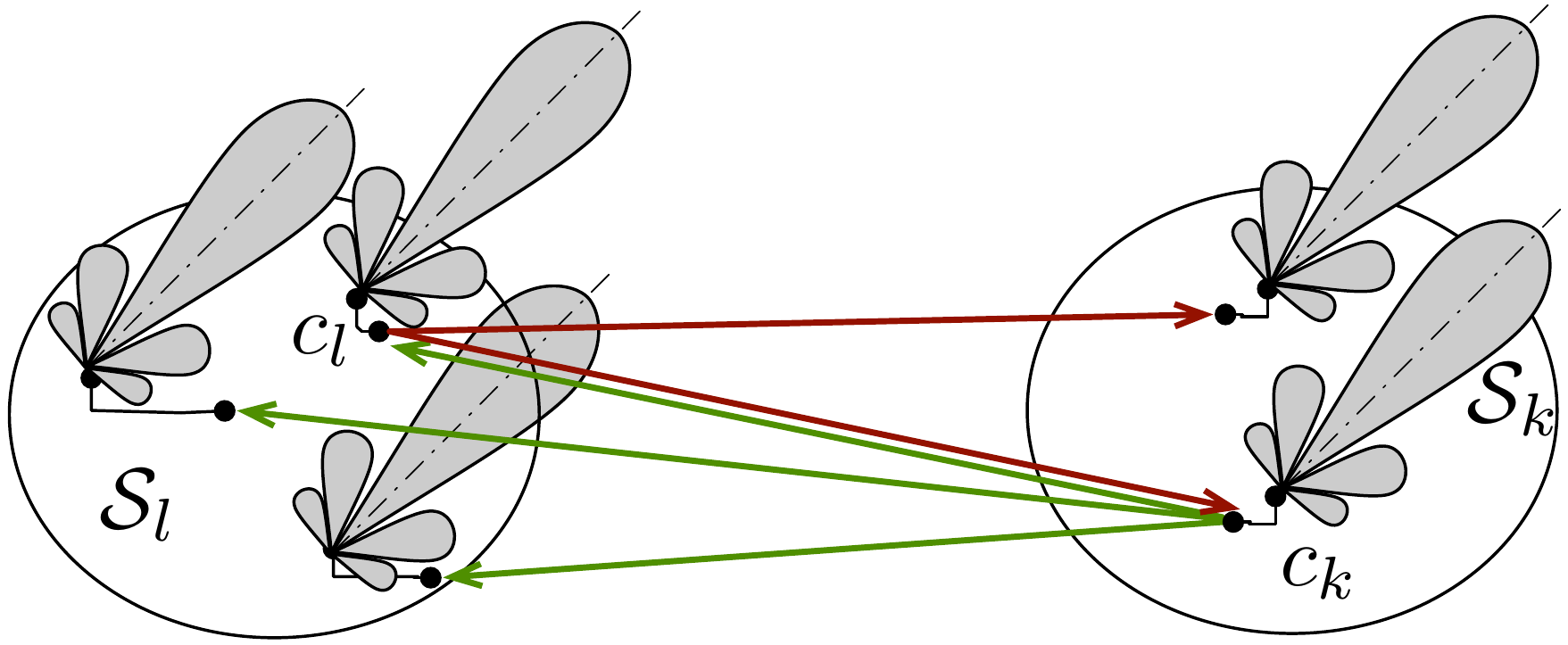}}
\caption{Reduced Aperture (RA) synthesis using super-stations $k$ and $l$.\label{parsynth}}
\end{figure}

In order to reduce the amount of transferred data, we propose to investigate a solution which consists in:
\begin{enumerate}
\item exploiting all the low-frequencies by computing locally all the
  correlations inside each super-station.
In this case, the low frequency term
associated to the spatial transfer function (denoted as $\mathcal{A}_R(\bu)$) is still $\mathcal{A}_L(\bu)$.
Denote as $\mathcal{S}_k$ the set of indices associated to
the beams in super-station $k$.
\begin{align*}
&\mathcal{A}_{l,l}(\bu) = \sum_{m<n\in \mathcal{S}_l} \delta(\bu-(\br_m-\br_n)) + \delta(\bu+(\br_m-\br_n))\\
& \mathcal{A}_L(\bu) =  \sum_{l} \mathcal{A}_{l,l}(\bu)
\end{align*}

\item Recovering the high frequency information by transferring only one single beam signal from each 
super-station to all other remote super-stations. 
If $c_k$ is the index of the reference beam in super-station $k$ transferred to
the other super-stations, the resulting sampling pattern of  visibilities associated to super-stations $k$ and $l$ is, see Fig. \ref{parsynth}:
\begin{align*}
 &\mathcal{A}_{k,l}(\bu) = \sum_{m \in \mathcal{S}_l} \delta(\bu-(\br_m-\br_{c_k}))\\
 &\mathcal{A}_H^R(\bu) = \sum_{k\not = l} \mathcal{A}_{k,l}(\bu)
\end{align*}
\end{enumerate}
This solution is motivated by the fact that it
allows to correctly sample the high frequencies,
at the cost of a reduced SNR, while reducing the number
of transfered electromagnetic signals. Other strategies than transferring a single antenna signal  are of
course possible. 
A solution which aims to preserve the SNR is to replace the signal indexed by $c_k$ by
averaging of neighboring beams. It is important to emphasize that unless the averaging process is 
taken into account in the measurement equation, 
the bias introduced by the averaging of nearby beams must be negligible.
This gain in SNR will be denoted as $\rho \geq 1 $ in the sequel.
Finally, the overall Reduced Aperture (RA) antenna spatial transfer function is:
\begin{equation}
\mathcal{A}_R(\bu) = \mathcal{A}_L(\bu) +  \rho \mathcal{A}_H^R(\bu)\label{reducedcover}
\end{equation}
This strategy reduces the transfer of beams data w.r.t. a centralized processing as long as the number of
stations inside each super-station is larger than the number of super-stations: for
$N_{ss}$ super-stations of $N_{s}$ stations each, the first requires $N_{ss}(N_{ss}-1)$ transfers
whereas the second  $N_{ss}N_{s}$.

\subsection{Analysis of performances on a simple model}

In order to evaluate analytically the loss related to the use of RA
synthesis \emph{w.r.t.} to a FA we consider a simple one dimensional case where the map is
a shifted Gaussian shape with flux $\alpha$:
\begin{equation}
x(p) = \frac{\alpha}{\eta \sqrt{2\pi}}e^{-\frac{(p-p_0)^2}{2\eta^2}}
\end{equation}
The unknown parameters are $\btheta=(\alpha,\eta,p_0)$.
The visibilities are:
\begin{equation}
v(u_\ell) = \alpha e^{\jmath 2\pi u_\ell p_0} e^{-2\pi^2 \eta^2 u_\ell^2}+ n_\ell,\; \ell=1\ldots M
\end{equation}
where $n_\ell$ is a measurement noise assumed independent Gaussian circular with
$n_\ell\sim \mathcal{N}_c(0,w_\ell^{-1}\sigma^2)$. The coefficient $w_\ell^{-1}$
takes into account the variance reduction that occurs when the visibility $v(u_\ell)$ 
is estimated from $w_\ell$ different baselines.

The elements of the  Fisher information matrix $I(\btheta)$ are computed using the Slepian-Bangs formula 
\cite[p. 293]{stoica2005spectral} which gives:
\begin{align}
&I(\btheta)  =\frac{2}{\sigma^2}\left(
\begin{array}{ccc}
 S_0 & -2 \pi^2 \alpha S_2 & 0 \\
 -2 \pi^2 \alpha S_2 & 4\pi^ 4 \alpha^2  S_4 & 0\\
 0 &0 & 4\pi^2\alpha^2 S_2
\end{array}
\right)\\
&S_q = \sum_{\ell=1}^M w_\ell u_\ell^q e^{-4\pi^2 \eta^2 u_\ell^2} 
\end{align} 
Note that $I(\btheta)$ is not a function of the source position $p_0$.

We compare the CRLBs on $\alpha$, $\eta$  and $p_0$ for two spatial
transfer functions.
In both cases the aperture configuration consists of two uniform sub-apertures separated by $D$
in order to sketch the behaviour of two super-stations:
\begin{align}
&r_k = -D/2 - k\Delta,\; k=0\ldots L\\
&r_{L+k+1} = D/2+k\Delta,\; k=0\ldots L
\end{align}
where $D > (L+1)\Delta $. In the FA mode, and for $u\geq0$:
\begin{align*}
&\mathcal{A}_L(u) = \sum_{\ell=0}^{L}2(L+1-\ell)\delta(u - \ell \Delta)\\
&\mathcal{A}_H^F(u) = \sum_{\ell=-L}^{L}(L+1-|\ell|)\delta(u - (D+(\ell+L) \Delta)) 
\end{align*}

\begin{figure}
\centering{\includegraphics[width=.6\columnwidth]{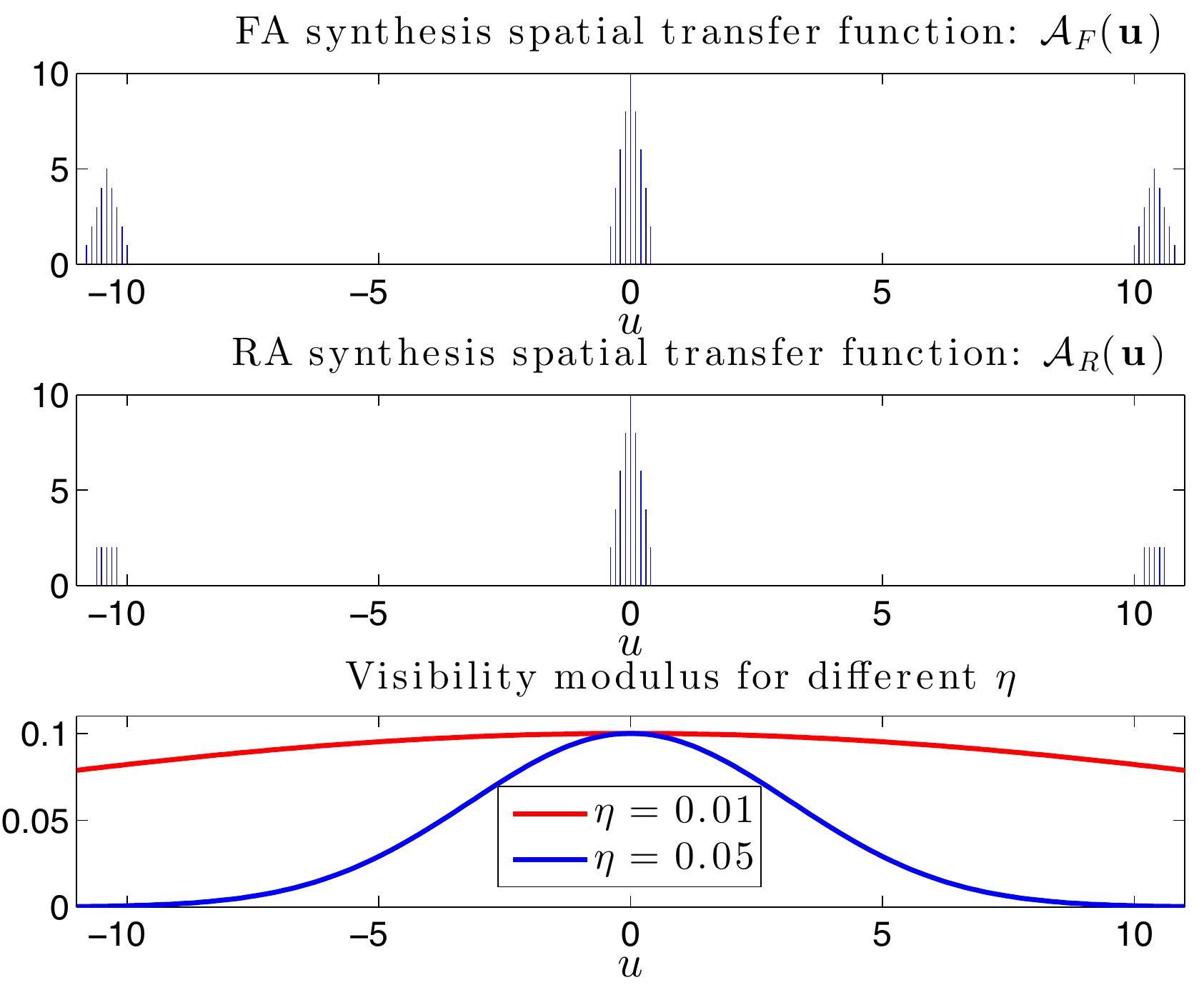}}
\caption{1D illustration of $\mathcal{A}_F(u)$ and $\mathcal{A}_R(u)$.
Bottom plot shows $|v(u)|$ for 2 characteristic values of $\eta$.
\label{CRLB1}}
\end{figure}

\begin{figure}
\centering{\includegraphics[width=.5\columnwidth]{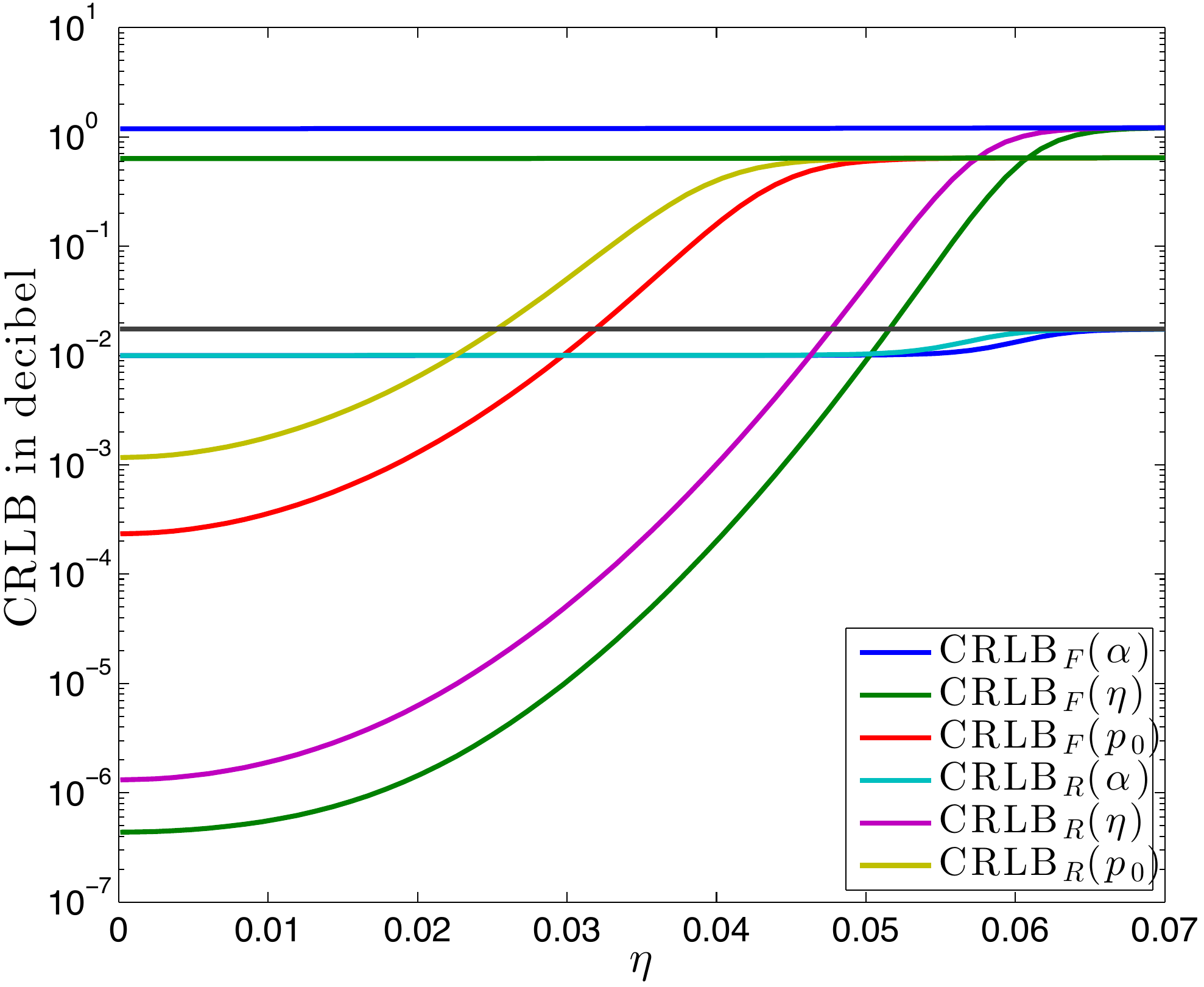}}
\caption{CRLB for the FA: $\text{CRLB}_F(\cdot)$, and the RA:
$\text{CRLB}_R(\cdot)$.\label{CRLB2}}
\end{figure}

$L$ is assumed even, $L = 2q$ and we consider in the
RA mode that $c_k=\pm q$: the reference beam is
in the middle of the opposite super-station. As noted above, the low frequency term $A_L(u)$ does not change.
The high frequency spatial transfer function is now for $u\geq 0$:
\begin{equation*}
\mathcal{A}_H^R(u)= \sum_{\ell=0}^{L} \delta(u - (D+(\ell+q) \Delta))
\end{equation*}
The source width $\eta$ plays a central role in the estimation. 
For a point source, $\eta \rightarrow 0$, high frequency measurements
will bring a lot of information  while performances
for a very extended source  ($\eta \rightarrow \infty$) will be independent of the inter-stations
visibilities. Figs.~\ref{CRLB1} and \ref{CRLB2} give the results obtained for a configuration 
defined by $L = 4$, $D = 10$ and $\Delta = 0.1$. The source parameters are $\alpha=1$, $p_0=0$ and results are 
given for different values of $\eta$. The gain $\rho$ is fixed to $\rho=2$.
Fig.~\ref{CRLB1} shows the two  spatial transfer functions
$\mathcal{A}_F(u)$ and $\mathcal{A}_R(u)$.
Fig.~\ref{CRLB2} shows the CRLBs associated to $A_F(u)$ and $A_R(u)$, denoted respectively as $\text{CRLB}_F$
and $\text{CRLB}_R$. The thresholding effect when $\eta\approx 0.05$ reflects the shape of the visibility
modulus given in Fig.~\ref{CRLB1}: the frequency contribution of the source at the inter-station baseline becomes negligible 
for $\eta>0.05$. For $\eta<0.05$ the order of magnitude of the loss of performances is 4dB.
This loss of performance naturally strongly depends on $\rho$, e.g. when $\rho=1$ the loss is 13dB.

\section{Distributed image reconstruction with partial aperture synthesis model}

\subsection{Decentralized map reconstruction}

The classical model for the map reconstruction is obtained
vectorizing the sampled map $\bx\in(\mathbb{R}^+)^N$ and the visibilities
$\bv \in \mathbb{C}^M$, $M<N$, and reads \cite{Giovannelli:2005is,Rau:2009ut}:
\begin{equation}
\bv =\bG\bx +\bn,\; \bG=\bW\bT\bF \label{reconstructFT} 
\end{equation}
where $\bF$ is the $N\times N$ Fourier transform matrix, $\bW$ is a diagonal 
weighting-matrix including various operations (calibration, signal to noise weighting),
and $\bT$ is a $0/1$ binary  $M\times N$  matrix which codes the sampling of the visibilities in the frequency plane.  The noise vector is assumed $\bn\sim \mathcal{N}_c(\boldsymbol{0},\sigma^2\bI)$.
With these assumptions, the image restoration problem is usually  casted as 
the general inverse problem:
\begin{equation}
  \label{CentInverseProblem}
  \hat{\bx}=\arg\min_{\bx\in (\mathbb{R}^+)^N} \|\bv - \bG\bx\|_2^2+\Omega(\bx)
\end{equation}
where the regularization term $\Omega(\bx)$ can be for example a sparse analysis  \cite{2012MNRAS.426.1223C},
a sparse synthesis \cite{2009MNRAS.395.1733W} or a hybrid prior \cite{Dabbech:2012ef}.

Denoting $\bT^\dagger  \bv$ the zero-padded visibilities and $\by = \bF^\dagger \bT^\dagger \bv$ the so called ``dirty image'', 
(\ref{reconstructFT}) can be rewritten as
\begin{equation}
\by = \bH \bx + \be
\end{equation}
where $\bH = \bF^\dagger \bT^\dagger \bW\bT\bF$ is a (circulant) convolution matrix.
Using $\forall \bz$ $\|\bT^\dagger \bz\|_2=\|\bz\|_2$ and $\bF^\dagger \bF =\bI$, 
we have $\|\bv - \bG\bx\|_2=\|\by - \bH\bx\|_2$ and
the inverse problem (\ref{CentInverseProblem}) turns to be equivalent to the deconvolution problem
\begin{equation}
  \label{CentDeconvProblem}
  \hat{\bx}=\arg\min_{\bx\in (\mathbb{R}^+)^N} \|\by - \bH\bx\|_2^2+\Omega(\bx)
\end{equation}

The vector $\bv$ can be partitioned as:
\begin{equation}
\bv^\dagger = (\bv_{1,\cdot}^\dagger,\bv_{2,\cdot}^\dagger,\ldots),\; 
\bv_{k,\cdot}^\dagger=(\bv_{k,1}^\dagger,\bv_{k,2}^\dagger,\ldots)
\end{equation}
where $\bv_{k,k}$ denotes the vector of visibilities obtained by correlating all the beams  inside super-station $k$ 
and vector $\bv_{k,l}$, $k\neq l $, by correlating beams in station $k$  with reference beam indexed by $c_l$ in station $l$.
More specifically $\bv_{k,k}$ is associated to $\mathcal{A}_{k,k}(\bu)$ and $\bv_{k,l}$ is associated to $\mathcal{A}_{k,l}(\bu)$, see sec. II.B. Using the same partitioning for $\bG$, $\bW$ and $\bT$, Eq~(\ref{CentInverseProblem}) can be rewritten as :
\begin{equation}
  \label{DecentInverseProblem}
  \hat{\bx}=\arg\min_{\bx\in (\mathbb{R}^+)^N} \sum_{k} \|\bv_{k,\cdot} - \bG_{k}\bx\|_2^2+\Omega(\bx)
\end{equation}
with $\bG_{k} = \bW_{k}\bT_{k}$. Using again the property of the zero padding matrix 
$\bT_{k,\cdot}^\dagger$ inside each norm in the sum (\ref{DecentInverseProblem}), we obtain:
\begin{equation}
  \label{DecentDeconvProblem}
  \hat{\bx}=\arg\min_{\bx\in (\mathbb{R}^+)^N} \sum_{k} \|\by_{k,\cdot} - \bH_{k}\bx\|_2^2+\Omega(\bx)
\end{equation}
where $\bH_{k} = \bF^\dagger \bT_{k}^\dagger \bW_{k}\bT_{k}\bF$. 
Note that $\bH = \sum_{k}\bH_{k}$ and $\by = \sum_{k}\by_{k}$.

In (\ref{DecentInverseProblem}), each sub-problem in the sum amounts to reconstruct $\bx$ from
the intra-super-station and inter-super-station visibilities $\bv_{k,\cdot}$.
In (\ref{DecentDeconvProblem}), $\by_{k,\cdot}$ is a dirty image obtained using only the visibilities
$\bv_{k,\cdot}$ and each sub-problem amounts to reconstruct $\bx$ from  $\by_{k,\cdot}$.

Eqs.  (\ref{DecentInverseProblem},\ref{DecentDeconvProblem}) are particularly interesting for the
derivation of distributed optimization algorithm, since they correspond to a
global variable consensus problem \cite{Boyd:2011tj,Sayed:2012uu}.
The next subsections evaluate 
the impact of the partial aperture models \textit{w.r.t.} the standard model
by numerical simulations.

\subsection{Array configuration and aperture synthesis}
\label{sec:tel_conf}

\begin{figure}
\centering{\includegraphics[width=.8\columnwidth]{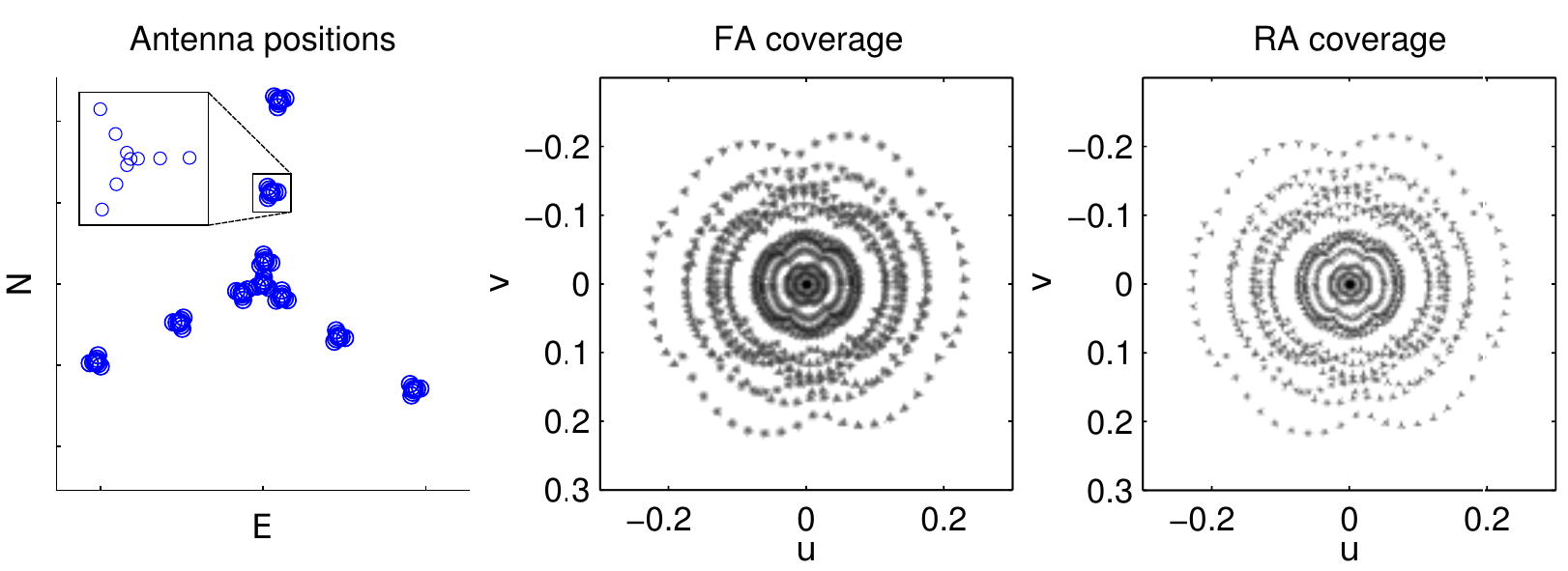}}
\caption{Left: sensor array configuration (without Earth rotation). Final $(u,v)$ coverage in normalized frequencies
for FA, Eq.~(\ref{fullcover}) and RA,  Eq.~(\ref{reducedcover}). 
\label{fig:array}}
\end{figure}

The shape of the sensor array used in the simulations consists
of 10 super-stations of 10 stations each, as shown in Fig.~\ref{fig:array}. The stations follow a
classical \textsf{Y} configuration with different rotations. 
Ten measurements are performed in a range of 3 hours taking into account
the Earth rotation. 
The RA is obtained by computing for each station the
correlation with the center beam of the other super-stations and with $\rho=2$.
For this experiment the ratio $D/(L\Delta)$ is $20$ and the visibilities are 
binned in a $256\times 256$ grid with Shannon sampling.
The corresponding $(u,v)$ coverage is  shown in  Fig.~\ref{fig:array}.

\begin{figure}
\centering{
\includegraphics[width=.75\columnwidth]{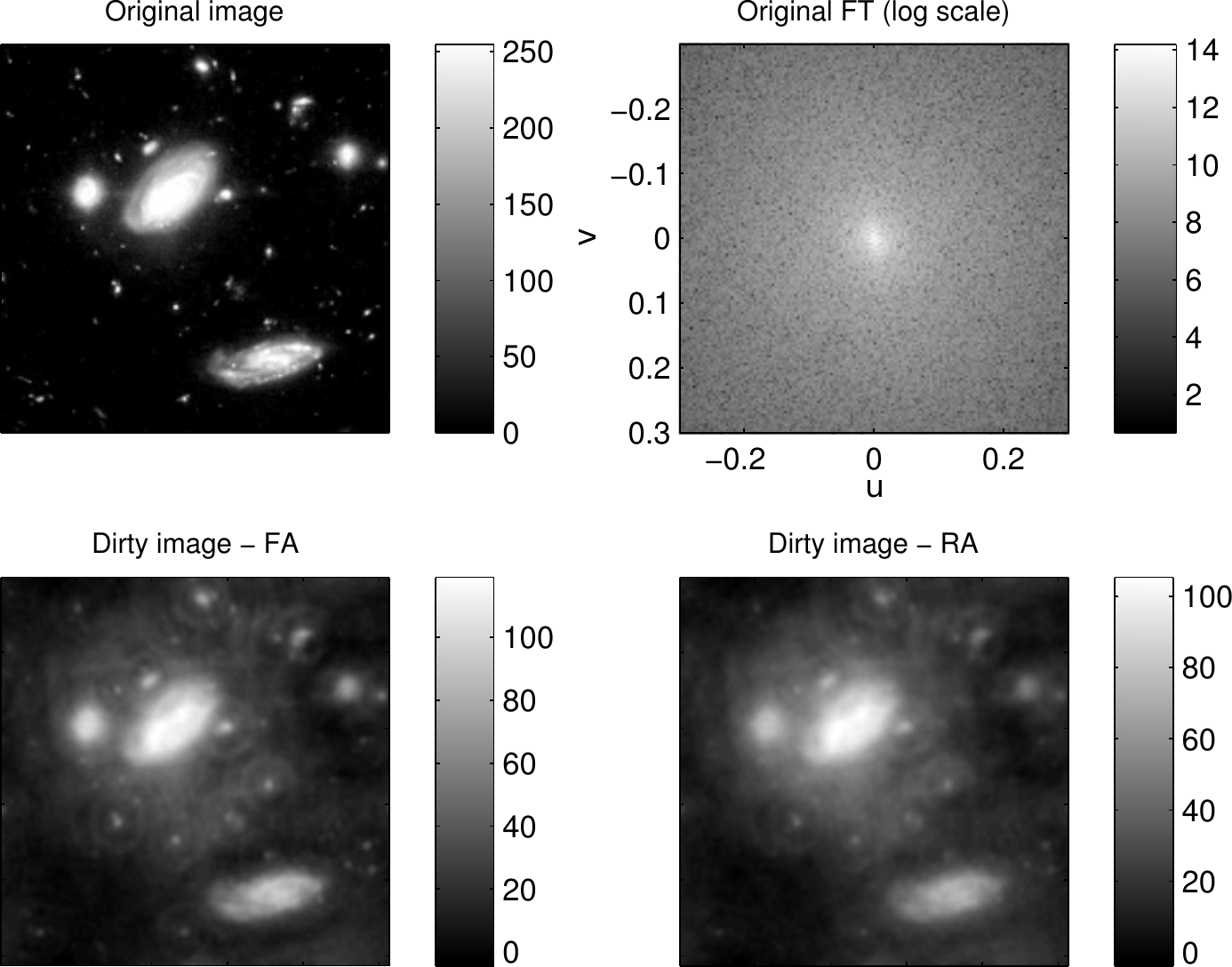}}
\caption{Original image and its Fourier transform (up). Dirty images
  observed with the full aperture (down left) and reduced aperture
  (down right). 
\label{fig:dirty}}
\end{figure}

The original image is shown in the
upper part of Fig.~\ref{fig:dirty} along with its Fourier
transform. Note that the image contains both low and
high frequency. The noise of variance $\sigma^2$ is such that the measured visibilities SNR is 70 dB. 
The observed dirty images  are also reported.  The RA clearly leads to a
less detailed dirty image.

\subsection{Distributed image reconstruction}
\label{sec:image_rec}

The image reconstruction is performed by  solving problem
(\ref{DecentDeconvProblem}) with a regularized  global variable consensus
ADMM algorithm as described in \cite{Boyd:2011tj}. 
The regularization term is
$\Omega(\bx)=\lambda\|\bx\|_2^2$ with $\lambda = 10^{-6}$. 
This regularization ensures that the problem is strictly convex
and limits the bias.
At each iteration, this algorithm requires to solve a large scale linear
problem of size the number of pixels $N$ in each super-station, \cite[Eq. (7.6)]{Boyd:2011tj}. 
This linear problem can be easily solved in the Fourier domain as discussed in
\cite{Giovannelli:2005is}. 
Note that the quadratic  regularization can be included in this step.
The consensus step  \cite[Eq. (7.7)]{Boyd:2011tj} is 
simply a projection on the positive orthant. A super-station sends the current image, Lagrangian multiplier and
receive the consensus image.

\begin{figure}
\centering{
\includegraphics[width=.75\columnwidth]{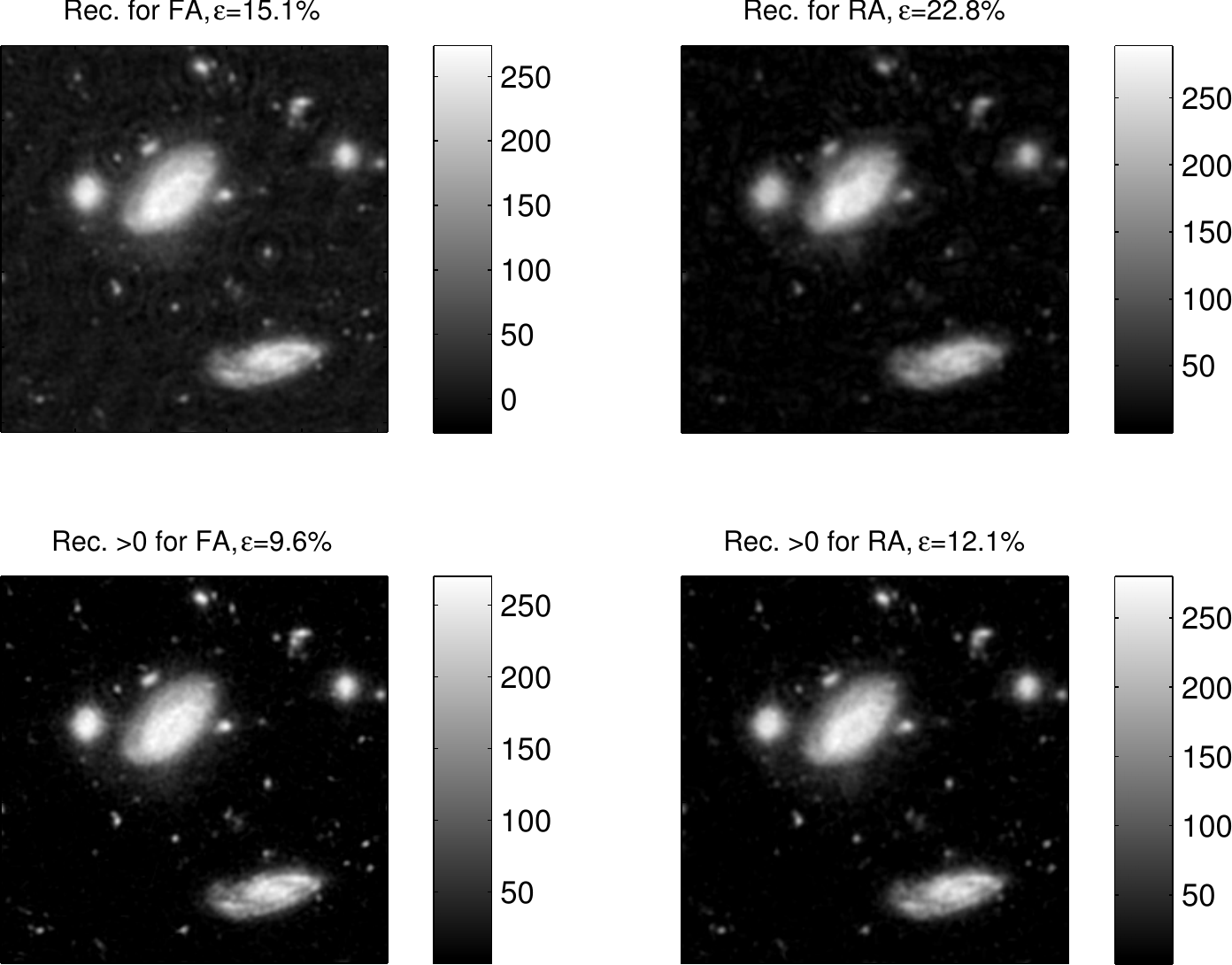}}
\caption{Reconstructed images and error $\epsilon$. Up (down): reconstruction
without (with) positivity constraint. Left : FA mode, right : RA mode. \label{fig:recdirty}}
\end{figure}

Fig.~\ref{fig:recdirty} shows the reconstructed images
obtained by solving (\ref{DecentDeconvProblem}) with
and without positivity constraints for both aperture cases, along with
the relative  norm of the error $\epsilon$.
The FA leads obviously to a better reconstruction in both cases.
However, while the loss in performance is rather important in the unconstrained case
($\epsilon = 15\%$ error instead of $\epsilon = 23\%$), this gap is significantly reduced
with the positivity constraint ($\epsilon = 9.5\%$ error instead of $\epsilon = 12\%$).

\section{Conclusion and perspectives}

This correspondence investigates a distributed strategy for
the image reconstruction problem in radio astronomy
when the number of stations inside each super-station 
is larger than the number of super-stations.
It relies on a reduced aperture synthesis where each super-station uses all its 
beam and a single reference beam from the other super-stations. 
Part of the missing information for each super-station is then exchanged during the  
consensus step of the distributed algorithm. 
The loss of performances, related to the use of a reduced aperture synthesis,
is evaluated on the image reconstruction  by computer simulations.

The approach proposed in the paper processes all the data after they have been acquired.
A natural extension is to reconstruct sequentially the image.
Among the benefits of this setup which perfectly fits the operating mode of interferometers
which progressively fill the frequency plane using the Earth rotation, is the possibility to 
optimize in real-time the observation mode. A straightforward solution is
to make a number of iterations of the reconstruction algorithm of section 
\ref{sec:image_rec} after each measurement and using a ``warm start''.
A much more challenging perspective it to select sequentially the 
frequency measurements used
at each iteration. Whereas this communication relies on a ``deterministic'' pattern of 
frequency measurements, a better strategy would be to select at each iteration, among all visibilities, 
the ones that optimize the reconstruction according to some predefined criterion.

\bibliographystyle{abbrv}
{
\tiny

}
\end{document}